\documentclass[12pt]{article}
\pdfoutput=1

\usepackage{latexsym,amsfonts,amsmath,amsthm,amssymb,amsbsy,multirow ,textcomp,hyperref,wrapfig,datetime,verbatim,cancel,subfigure,cite}
\usepackage[dvipdfmx]{graphicx}
\usepackage[font={footnotesize,sl},bf]{caption}

\def\be{\begin{equation}}
\def\ee{\end{equation}}
\def\bea{\begin{eqnarray}}
\def\eea{\end{eqnarray}}
\newcommand{\beq}{\begin{eqnarray}}
\newcommand{\eeq}{\end{eqnarray}}

\numberwithin{equation}{section}

\long\def\new#1\endnew{{\bf #1}}		
\long\def\del#1\enddel{}

\def\del{\partial}
\def\nn{\nonumber}

\usepackage{color}
\usepackage[dvipsnames]{xcolor}

\newcommand{\pink}[1]{\textcolor{\pink}{#1}}

\def\nn{\nonumber}

\def\tg{\widetilde \gamma}
\def\talpha{\widetilde \alpha}
\def\tpsi{\widetilde \psi}
\def\tphi{\widetilde \phi}

\newcommand{\ket}[1]{\left| #1\right>}

\begin{document}

 \begin{titlepage}
  \thispagestyle{empty}
  \begin{flushright}
  \end{flushright}
  \bigskip
  \begin{center}
  \baselineskip=13pt {\LARGE A rough end for smooth microstate geometries}
   \vskip1.5cm
   \centerline{
   {\bf Donald Marolf}\footnote{\tt marolf@physics.ucsb.edu}${}^\blacklozenge$,
   {\bf Ben Michel}\footnote{\tt michel@physics.ucsb.edu}$^\blacklozenge$,
   {\bf Andrea Puhm}\footnote{\tt puhma@physics.harvard.edu}${}^\diamondsuit{}^\odot$
   }
  \bigskip
    \bigskip
   \centerline{\em${}^\blacklozenge$ Department of Physics, University of California}
 \centerline{\em Santa Barbara, CA 93106 USA}
   \bigskip
 \centerline{\em${}^\diamondsuit$ Jefferson Physical Laboratory, Harvard University}
 \centerline{\em Cambridge, MA 02138, USA}
 \bigskip
  \centerline{\em${}^\odot$ Black Hole Initiative, Harvard University}
  \centerline{\em Cambridge, MA 02138, USA}

  \vskip2cm
  \end{center}

  \begin{abstract}
  \noindent
Supersymmetric microstate geometries with five non-compact dimensions have
recently been shown by Eperon, Reall, and Santos (ERS) to exhibit a
non-linear instability featuring the growth of excitations at an
``evanescent ergosurface'' of infinite redshift.  We argue that this growth
may be treated as adiabatic evolution along a family of exactly
supersymmetric solutions in the limit where the excitations are
Aichelburg-Sexl-like shockwaves.  In the 2-charge system such solutions may be constructed
explicitly, incorporating full backreaction, and are in fact special cases
of known microstate geometries.  In a near-horizon limit, they reduce to
Aichelburg-Sexl shockwaves in $AdS_3 \times S^3$ propagating along one of
the angular directions of the sphere.  Noting that the ERS analysis is valid
in the limit of large microstate angular momentum $j$, we use the above
identification to interpret their instability as a transition from rare
smooth microstates with large angular momentum to more typical microstates with
smaller angular momentum.  This entropic driving terminates when the angular
momentum decreases to $j \sim \sqrt{n_1n_5}$ where the density of
microstates is maximal.  We argue that, at this point, the large stringy corrections
to such microstates will render them non-linearly stable.  We identify a possible mechanism for this stabilization and detail an
illustrative toy model.
\end{abstract}

 \end{titlepage}

\setcounter{tocdepth}{2}

\tableofcontents

\section{Introduction}

The black hole information paradox \cite{Hawking:1974sw,Mathur:2009hf} and more recently the firewall argument \cite{Almheiri:2012rt,Almheiri:2013hfa} have reignited the search for the correct microscopic description of black holes.  The study of supersymmetric black holes in string theory has been a useful arena for this study, providing many insights.  For example, such black holes may be described as bound states of strings and branes \cite{Callan1996}, which can then be explored using either the low-energy perturbative worldvolume gauge theory on the branes or supergravity  at finite coupling \cite{Horowitz:1996nw}. One of the great triumphs of this approach is the explicit stringy counting~\cite{Strominger:1996sh} by Strominger and Vafa of the number of microstates of the D1-D5-P system, which famously agrees precisely with the Bekenstein-Hawking entropy of the naive black hole solution.

The fuzzball program~\cite{Mathur:2005zp,Bena:2007kg,Balasubramanian:2008da,Skenderis:2008qn,Mathur:2008nj,Chowdhury:2010ct,Bena:2013dka} is an attempt to describe these microstates at finite coupling.  It argues that the extended objects of string theory modify the structure of the black hole horizon and solves the information paradox by construction: there is no horizon, only an end to spacetime. Some of the major goals of the program are to explain the Bekenstein-Hawing entropy, construct representative microstates and, especially in light of the firewall paradox, to understand the consequences of the stringy/braney physics at the horizon.

Within this program one may distinguish 3 types of microstates ~\cite{Bena:2013dka}: (i) microstate {\it geometries}, smooth horizonless solutions of supergravity; (ii) microstate {\it solutions}, horizonless solutions of supergravity with singularities corresponding to D-brane sources or which can be dualized patch-wise into smooth geometries; and (iii) general {\it fuzzballs}, horizonless configurations which may be arbitrarily quantum and/or strongly curved. Since the horizon is a classical notion, it may well be that this definition of general {\it fuzzball} includes all black hole microstates in any approach to the information problem.  In any case, it remains an open question what fraction of black hole microstates fall into each category. In particular, while in several examples of supersymmetric black holes it has been argued \cite{Lunin:2002iz,Bena:2013dka} that many microstates do in fact have a consistent description entirely within supergravity, it is far from clear that they are typical.

New questions about this program were recently raised by Eperon, Reall and Santos (ERS) \cite{Eperon:2016cdd}.  Focusing on supersymmetric microstate {\it geometries}, they identified a non-linear classical instability due to the growth of excitations at an ``evanescent ergosurface''~\cite{Gibbons:2013tqa} of infinite redshift. On such a surface, there are null geodesics with zero energy relative to infinity which are {\it stably trapped} in the potential well near the ergosurface. They find that perturbing the microstate by adding a massive particle or general wavepacket near the evanescent ergosurface eventually leads to large backreaction, even if the particle has negligible energy at infinity.  In particular, the coupling of the particle to supergravity fields will allow it to gradually radiate energy and angular momentum and its trajectory will approach a geodesic that minimizes the energy. Since the particle is now following an almost-null trajectory, the local energy and hence backreaction will be very large.
The instability is non-linear in the sense that it involves interactions between the particle and the radiation field. A corresponding effect arises in perturbative field theory due to the coupling of modes near the evanescent ergosurface (playing the role of the massive particle above) to radiative degrees of freedom at infinity.~\footnote{It has long been known that a class of non-supersymmetric fuzzball solutions \cite{Jejjala:2005yu} exhibits a linear ergoregion instability~\cite{Cardoso:2005gj}. However, such a stability analysis of supersymmetric microstate geometries had not been performed until the recent work by ERS. Another recent study of dynamics focuses on the quantum tunneling of branes into microstate geometries \cite{Bena:2015dpt}; the result suggests that a collapsing shell of matter might tunnel into a fuzzball configuration before a horizon can form.}

The emission of angular momentum reduces the size of a fuzzball.  However, at least in well-understood cases, typical fuzzballs have structure on microscopic scales and thus are not described by smooth solutions \cite{Chen:2014loa}. The ERS instability implies that smooth solutions can only describe the system for a short time when it is coupled to the environment.
In a dual CFT description of the near-horizon region,  the instability corresponds to motion among the ground states towards larger (and more generic) twist numbers~\cite{Dijkgraaf:1998gf,Seiberg:1999xz}.  As a result, and as we emphasize below, such an instability might have been deduced on entropic grounds even before the identification of a dynamical mechanism by ERS.

The implications of the ERS instability for the fuzzball program depend on its endpoint.   ERS proposed that it could lead to a collapse of the evanescent ergosurface and thus drive the initially smooth horizonless microstate geometry to an almost-supersymmetric black hole with the same brane charges as the microstate geometry but  with different angular momenta. In particular, they suggested that the endstate of the instability (for supersymmetric D1-D5 microstates with additional momentum charge) might be a near-extremal black hole~\cite{Breckenridge:1996is} or a black ring\cite{Elvang:2004rt}.
To support this argument one may note that as the solution shrinks it is described by the duality cascade of~\cite{Chen:2014loa}, but since the evanescent ergosurface is a consequence of supersymmetry it persists in every duality frame and so the ERS instability argument continues to apply.

However, entropic reasoning leads to the expectation that the endpoint is instead a typical microstate with angular momentum $j_\text{typical}$ which maximizes the microstate density of states $S(j)$.  In particular, we suggest that the string-scale structure of a typical microstate leads to corrections that remove the instability for $j \sim j_\text{typical}$ and prevents the collapse to a black hole.  Within the supergravity approximation the stabilized geometry is indistinguishable from a supergravity black hole but has structure at the horizon that differentiates the two in the full string theory.  This structure is located at the bottom of the duality cascade described in~\cite{Chen:2014loa}, and supergravity will not capture the full physics at the fuzzball core.

To obtain a measure of analytic control over the ERS instability,  we take an adiabatic limit in which the particle is well-described by an  Aichelburg-Sexl-like shockwave on the evanescent ergosurface. We focus on 2-charge microstates, for which the general microstate geometries are known. Solutions with such shockwaves preserve the same supersymmetries as the microstate geometries and are thus independent of time, but a small departure from this limit will lead to slow evolution.  In particular, growth of the instability leads to growth of the shockwave and thus to motion along this family of solutions.  The geometries accounting for the backreaction of the shock are known explicitly \cite{Lunin:2002fw} and in fact correspond to special cases of the more general family of microstate geometries. The CFT states dual to their near-horizon limits were described in \cite{Lunin:2002bj}.  These facts can be used to justify the entropic reasoning used above.

Analysis of any potential instability in typical microstates would require a better understanding of black hole microstates beyond supergravity. In the absence of such knowlege, we describe a simple toy model displaying what we believe to be key features of their stringy physics.  In particular, the model includes both a low-energy region near the evanescent ergosurface, a parameter that we also call $j$ controlling the microstate size, and an analog of the internal structure that would be associated with stringy excitations used to perturb the microstates.   We then study the model as one decreases $j$ in analogy with the adiabatic evolution described above.  At small enough $j$ the low-energy region displays features on scales smaller than those set by the internal structure of the probe.  The probe can then no longer take full advantage of the low-energy region, raising the ground state energy and shutting off the instability.  Thus we argue that the net effect of the ERS instability is to drive smooth solutions through the duality cascade of \cite{Chen:2014loa} towards typicality, and the instability is stabilized by stringy corrections just as supergravity breaks down: a rough end for smooth microstate geometries.

The organization of this paper is as follows. In~\S\ref{sec:2charge} we review some of the salient features of the supergravity and CFT descriptions of the 2-charge system. We then address the ERS instability in~\S\ref{sec:instability}. After reviewing the main argument of~\cite{Eperon:2016cdd},  we study Aichelburg-Sexl-like shockwaves described above and discuss their identification in terms of known microstate geometries. This allows us to give a concrete description of adiabatic evolution along this family.  \S\ref{ssec:qmmodel} then describes and analyzes our toy model illustrating our proposed mechanism for stabilizing the system once the microstates become typical.    We conclude with a discussion of our results  in~\S\ref{sec:conclusions}. Appendix ~\S\ref{app:3charge} describes the analogous physics for a special class of 3-charge microstate solutions.

\section{2-charge microstates}\label{sec:2charge}

Our analysis will focus on 2-charge supersymmetric microstate geometries; discussion of the 3-charge case is relegated to appendix \ref{app:3charge}.
There is now considerable evidence~\cite{Lunin:2001jy,Lunin:2002qf,Mathur:2005zp} supporting the identification of particular states $|\Psi\rangle$ in the D1-D5 CFT at small string coupling $g_s$ and large brane charges $Q_1,Q_5$ with (the near-horizon limit of) a class of horizonless supergravity solutions characterized by a profile $\vec{F}$ in the four non-compact transverse spatial dimensions.  The map between these descriptions takes the form
\begin{equation}\label{eq:CFTgeom}
 |\Psi\rangle=\prod_{k=1}^{N} (\sigma_k^{ss'})^{N_k}\ket{0} \quad \longleftrightarrow \quad \vec{F}(v) = \sum_{k=1}^N \vec{F}_k e^{ik\omega v}\,,
\end{equation}
where the $N_k$ are related to the Fourier amplitudes $\vec{F}_k$.
We will discuss the details of the CFT and supergravity descriptions, and thus the two sides of \eqref{eq:CFTgeom},  in ~\S\ref{sec:cft} and ~\S\ref{sec:geometry}.

\subsection{CFT Review} \label{sec:cft}

Let us consider IIB string theory compactified to $M^{1,4}\times S^1\times T^4$, with $n_1$ D1 branes wrapping the $S^1$ and $n_5$ D5 branes wrapping $S^1\times T^4$. At parametrically large $S^1$ the low-energy dynamics of the bound state of these branes is described by a $(1+1)$ dimensional sigma model whose target space is the moduli space of $n_1$ instantons in the D5-brane gauge theory~\cite{Dijkgraaf:1998gf,Seiberg:1999xz}, a resolution of the orbifold ${(T^4)}^N /S_N$ (the symmetric product of $N = n_1 n_5$ copies of $T^4$). The CFT has $\mathcal{N} = (4,4)$ supersymmetry and a moduli space of supersymmetric deformations. It is conjectured that this moduli space contains the ``orbifold point'' where the target space is just the orbifold ${(T^4)}^N/S_N$. This is the symmetric product of a seed with 4 real bosons $X_i$ (4 torus directions), 4 real left moving fermions $\psi_i$, 4 real right-moving fermions $\psi_i'$ and central charge $c=6$.

The complete theory with target space ${(T^4)}^N/S_N$ has $N$ copies of the $c=6$ CFT with states symmetrized between the $N$ copies. Many details of this theory are given in~\cite{Balasubramanian:2005qu}, here we just review some relevant aspects. Modular invariance requires that we introduce twisted sectors, created by bosonic and fermionic twist operators permuting the $N$ copies. These operators are labeled by conjugacy classes of cycles of $S_N$, which can be decomposed into irreps $\sigma_k$ labeled by a single cycle of length $k$ (the particular elements are irrelevant because of the symmetrization, which will be implicit). For simplicity, in our discussion below we place all oscillators ossciated with the $T^4$ in their ground state.   A general such twisted sector state corresponds to
\begin{equation}\label{eq:twist}
\ket{\{N_k\}}=\prod_{k=1}^{N} (\sigma_k^{ss'})^{N_k}\ket{0}
\end{equation}
where $s,s'=\pm$ and
\begin{equation}
\label{eq:levelcondition}
\sum_{k=1}^N k N_k = N\,,
\end{equation}
since each copy must be involved in the permutation.
We take the field theory on the D1-D5 system to be in the Ramond sector \cite{Lunin:2002iz}.
The $\sigma_k$ in~\eqref{eq:twist} have $(h,\tilde{h})=(\frac{c}{24},\frac{c}{24})$ and so any set of $\{N_k\}$ satisfying~\eqref{eq:levelcondition} is a Ramond ground state. This fact underlies the argument matching the ground state degeneracy with the black hole entropy.

In the D1-D5 CFT the R-symmetry is geometrized as the rotational symmetry of the non-compact directions $SO(4) \approx SU(2)_L\times SU(2)_R$. Maximal R-charge hence corresponds to maximal angular momentum. The left-moving fermions $\psi_i$ carry spin $\frac{s}{2}$ under $SU(2)_L$ while the right-moving fermions $\psi_i'$ carry spin $\frac{{s'}}{2}$ under $SU(2)_R$; the R-charge of the state is given by $(j,j')=(j_L^3,j_R^3)$.
The $\sigma_k$ form bi-doublets of the $SU(2)\times SU(2)$ R-symmetry and in~\eqref{eq:twist} and below we take the $(s,{s'})=(-,-)$ component.

We can now explain the main features of the density of states $S(j)$ as a function of angular momentum.  The state $\ket{N_1}=\sigma_1^{N_1}|0\rangle$ with $N_1=N$ and all other modes zero is the unique completely untwisted state, corresponding to the Ramond ground state with maximal R-charge, and thus to a state of maximal angular momentum $j_\text{max}=n_1n_5$.   Less finely-tuned states have smaller angular momentum, so $S(j)$ is a decreasing function of $j$ near $j_\text{max}$. Indeed, for $j \gg \sqrt{n_1n_5}$ (and once the oscillators associated with the internal $T^4$ are included as well) one finds \cite{Balasubramanian:2005qu}
\begin{equation}
\label{eq:Slargej}
S(j) = 2\pi\sqrt{2} \sqrt{n_1n_5-|j|}
\end{equation}
to leading order in $N$. On the other hand, since the twist operators can contribute angular momentum with any sign, charge conjugation symmetry implies that the ensemble of all ground states has vanishing expectation value for the angular momentum.  Fluctuations about the average imply typical states to have non-zero angular momentum of order $\sqrt{N} = \sqrt{n_1n_5}$, so $S(j)$ is maximized in this regime and decreases when $j$ is decreased further.

Before proceeding to discuss geometries, we remind the reader that states in the Ramond sector can be mapped to states in the Neveu-Schwarz sector via a symmetry of theories with $\mathcal{N}\geq 2$ in 2 dimensions known as spectral flow. The dimensions $h$ and R-charges $j$ of operators change along the flow according to\cite{Schwimmer:1986mf}:
\begin{equation}
\label{eq:CFTSF}
 h_\alpha=h-\alpha j+\alpha^2 \frac{c}{24}\,,\quad\quad
 j_\alpha=j-\alpha \frac{c}{12}\,.
\end{equation}
In particular, a Ramond ground state with maximal R-charge $(h,j)=(\frac{c}{24},\frac{c}{12})$ can be mapped via~\eqref{eq:CFTSF} with $\alpha=1$ to the Neveu-Schwarz vacuum $(h,j)=(0,0)$. Ramond ground states of non-maximal R-charge map to chiral primaries in the NS sector.  As a result, the completely untwisted state $|N_1\rangle$ becomes the NS vacuum dual to global $AdS$.  In particular, on the gravity side spectral flow of the near-horizon limit for the corresponding solution will give simply $AdS_3 \times S^3$ in global coordinates.

\subsection{Geometries}\label{sec:geometry}

The two-charge D1-D5 geometries are type IIB compactifications on $S^1\times T^4$ (or K3) characterized by a curve $\vec{F}(v)$ in $\mathbb{R}^4\times T^4$.  Due to the fact that these solutions were originally constructed in a duality frame where the charges are P-F1, the curve $\vec{F}(v)$ is known as the string profile.

We will focus on solutions describing only oscillations in the four non-compact transverse directions $x$. The complete solution with oscillations in the $T^4$ directions $z$ is given in Refs.~\cite{Lunin:2002iz,Kanitscheider:2007wq}. Since the $T^4$ factor plays no further role in our discussion of the ERS instability we will usually omit it henceforth. The argument $v=t-y$ of the string profile is a lightcone coordinate involving the spatial coordinate $y$ along the $S^1$. The metric, dilaton and RR 2-form for such solutions are given by~\cite{Lunin:2002iz}
\begin{eqnarray}
ds^2 &=& \frac{1}{\sqrt{H_1 H_5}}\left[-(dt-A)^2+(dy+B)^2\right]+\sqrt{H_1H_5}dx_4^2+\sqrt{\frac{H_1}{H_5}}\sqrt{V}dz_4^2
\,,\nonumber\\
e^{\Phi} &=&g\sqrt{\frac{H_1}{H_5}}
\,,\nonumber\\
C_2 &=& g^{-1} \left[ H_1^{-1} (dt-A)\wedge (dy+B) +\zeta\right]
\,, \label{eq:fuzzgeos}
\end{eqnarray}
where the harmonic functions are
\begin{equation}
\label{d1d15h}
H_5 = 1 + \frac{Q_5}{L}\int_0^L \frac{dv}{|\vec{x}-\vec{F}(v)|^2}
\,, \;
H_1 = 1 + \frac{Q_5}{L}\int_0^L \frac{|\dot{\vec{F}}|^2dv}{|\vec{x}-\vec{F}(v)|^2}
\,,\;
A^i = -\frac{Q_5}{L}\int_0^L\frac{\dot F^i dv}{|\vec{x}-\vec{F}(v)|^2} \,.
\end{equation}
The remaining quantities are defined via $dB = \star_4 dA$, $d\zeta = -\star_4 dH_5$.\footnote{Our functions $H_5,H_1,A$ correspond, respectively, to $H^{-1},K+1,A$ in {\it e.g.}~\cite{Lunin:2002bj}.} $L= \frac{2\pi Q_5}{R}$, and its presence in \eqref{d1d15h} is a vestige of the original derivation of these solutions. The profile $\vec{F}$ relates the D5 charge $Q_5$ to the D1 charge:
\begin{equation}\label{eq:Q1}
 Q_1=\frac{Q_5}{L} \int_0^L |\dot{\vec{F}}|^2dv\,.
\end{equation}
These supergravity charges $Q_1,Q_5$ are related to the dimensionless quantized charges $n_1,n_5$ by
\begin{equation}\label{eq:n1n5}
 Q_1=\frac{g \alpha'^3}{V}n_1\,, \quad Q_5=g \alpha' n_5\,.
\end{equation}
The $y$ coordinate is identified under $y\rightarrow y+2\pi R$ and $V$ is the asymptotic volume of the $T^4$ whose coordinates $z$ have period $2\pi$. The four flat transverese directions $x$ are non-compact and can be coordinatized as $dx_4^2=d\widetilde r^2+\widetilde r^2(d\widetilde \theta^2+\sin^2\widetilde \theta d\widetilde \phi^2+\cos^2\widetilde \theta d\widetilde \psi^2)$. The relation between the Cartesian coordinates $(x_1,x_2,x_3,x_4)$ and the spherical coordinates $(\widetilde r, \widetilde \theta, \widetilde \phi, \widetilde \psi)$ is given by $x_1=\widetilde r \sin \widetilde \theta \cos \tphi$,  $x_2=\widetilde r \sin \widetilde \theta \sin \tphi$, $x_3=\widetilde r \cos \widetilde \theta \cos \tpsi$,  $x_4=\widetilde r \cos \widetilde \theta \sin \tpsi$.

Supersymmetry fixes the energy to be
\begin{equation}
\label{eq:energy}
 \mathcal{E}=Q_1+Q_5\,
 \end{equation}
while the angular momentum depends on $\vec{F}$ through \cite{Lunin:2002iz}
\begin{equation}\label{eq:J}
J_{ij}=\frac{Q_5 R}{L}\int_0^L(F_i \dot{F}_j-F_j \dot{F}_i)dv\,.
\end{equation}
This quantity has dimensions $[\text{length}]^4$ and is related to the quantized angular momentum $j$ by
\be
J_{12} = \frac{g^2}{V} j,\quad\quad J_{34} =  \frac{g^2}{V} j'
\ee
in units where $\alpha'=1$. For details relevant to computing energy and angular momentum in the above 6d geometries, see~\cite{Townsend:2001rg,deLange:2015gca}.

It will be useful to estimate the size of a given curve $\vec{F}$ as this determines the validity of the supergravity description at the string profile \cite{Chen:2014loa}.   As argued in \cite{Balasubramanian:2005qu}, the size of the curve is roughly proportional to its angular momentum $J = \sqrt{J_{ij} J^{ij}}$.
The $\vec{F}$ that carries maximal angular momentum $J_\text{max}=Q_1 Q_5$ extends to a distance $\sqrt{Q_1 Q_5}/R$ from the center while strings carrying a fraction $J_\text{max}/m$ of the maximum angular momentum are smaller by a factor $1/m$.  As noted in \S~\ref{sec:cft},  most CFT states have $j_\text{typical}/j_{\rm max} \approx 1/\sqrt{n_1 n_5}$ and so have size of order $1$ in string units. The supergravity description is valid (i.e. weakly curved) in the large $N=n_1 n_5$ limit, so from this perspective both $J_\text{typical}$ and the typical size are indistinguishable from zero. Indeed, in the strict supergravity limit one may compute the density of states in direct analogy with \cite{Palmer:2004gu,Bak:2004rj,Bak:2004kz,Rychkov:2005ji} to obtain \eqref{eq:Slargej} (which is maximized only at $j=0$).

In a different duality frame the solution~\eqref{eq:fuzzgeos} describes a singular string source along $\vec{F}(v)$ carrying momentum, but in the corner of moduli space where the asymptotic charges are D1-D5 it has long been argued \cite{Lunin:2002iz} that the geometry is completely smooth. This feature is particularly intriguing, as the ensemble of 2-charge solutions approximates the $M=0$ BTZ black hole~\cite{Balasubramanian:2005qu}, so one could argue that the actual black hole microstates were horizon-free geometries that cap off smoothly at the string profile.  However, for typical states it turns out \cite{Chen:2014loa} that maintaining the validity of the supergravity description while descending toward the fuzzball requires a duality cascade.  Furthermore, the cascade terminates in a frame where the D1-D5 charges have become P-F1 and curvature of the $S^3$ becomes string-scale, so that even this final supergravity description breaks down near the location of the typical string profile.  Typical 2-charge states are thus not well-described by smooth geometries. However, states with atypically large angular momenta have string profiles that vary slowly enough for supergravity to remain valid even at the locus defined by $\vec{F}(v)$, in some cases using only a single duality frame. Such states are indeed described by smooth geometries.

It is therefore of particular interest that ERS~\cite{Eperon:2016cdd}  found an instability for the geometry with maximal angular momentum which is the prime example of such a solution. Since we will also begin our discussion of shockwaves in \S\ref{sec:instability} with this special case, we now pause to describe it in some detail.

\subsubsection{The maximally-rotating microstate}
\label{smoothgeom}

The angular momentum \eqref{eq:J} obtains its maximum value for the profile function
\begin{eqnarray}\label{eq:maxrotF}
 \vec{F}(v) = (a \cos (\omega v)\,, a \sin (\omega v),0,0)\,, \quad 0\leq v \leq L\,,
\end{eqnarray}
where
\begin{equation}
 a=\frac{\sqrt{Q_1 Q_5}}{R}\,, \quad \omega = \frac{2\pi}{L}\,.
\end{equation}
The D1 charge~\eqref{eq:Q1} for this profile is
\begin{equation}
 Q_1=Q_5 a^2 \omega^2\,,
\end{equation}
and the angular momentum~\eqref{eq:J} in the $x_1-x_2$ plane, or equivalently, along the $\widetilde \phi$ direction, takes the value
\begin{equation}
 J_{\widetilde \phi}= J_{12}=Q_1 Q_5=J_\text{max}\,.
\end{equation}
With the profile~\eqref{eq:maxrotF} the harmonic functions become (in the notation of~\cite{Mathur:2005zp})\footnote{We use coordinates ($r,\theta,\tphi,\tpsi$) in which the flat metric takes the form
\begin{equation}
 dx_4^2 = (r^2+a^2\cos^2\theta)\left(\frac{dr^2}{r^2+a^2} + d\theta^2\right) + (r^2+a^2) \sin^2\theta d\tphi^2 + r^2 \cos^2 \theta d\widetilde  \psi^2\,,
\end{equation}
and $\epsilon_{r\theta \tphi \tpsi}=\sqrt{g}=(r^2+a^2\cos^2\theta) r \sin\theta \cos\theta$. These coordinates are related to $(\widetilde r, \widetilde \theta, \widetilde \phi, \widetilde \psi)$ in which the $S^3$ takes its standard form $d\Omega_3^2= (Q_1 Q_5)^{1/4} (d\widetilde \theta^2 + \sin^2 \widetilde \theta d\tphi^2 + \cos^2\widetilde \theta d\tpsi^2)$ by $ \widetilde r = \sqrt{r^2+a^2 \sin^2\theta}$ and $\cos \widetilde \theta = \frac{r\cos\theta}{\sqrt{r^2+a^2\sin^2\theta}}$.
}
\begin{align}
&H_5= 1+\frac{Q_5}{r^2+a^2 \cos^2\theta}\,, \quad H_1= 1+\frac{Q_1}{r^2+a^2 \cos^2\theta}\,,\\
&A_{\tphi}=-Q_5 a^2 \omega \frac{\sin^2 \theta}{(r^2+a^2\cos^2\theta)}\,,\quad B_{\tpsi}=-Q_5 a^2 \omega \frac{\cos^2 \theta}{(r^2+a^2\cos^2\theta)}\,.
\end{align}
The full solution is given by
\begin{eqnarray}\label{eq:metricR}
 ds^2_R&=&-\frac{1}{h} (dt^2-dy^2) + hf \left(d\theta^2+\frac{dr^2}{r^2+a^2} \right) -\frac{2a \sqrt{Q_1 Q_5}}{hf} \left(\cos^2\theta dy d\tpsi + \sin^2\theta dt d\tphi \right)\nn\\
 &&+h\left[\left(r^2+\frac{a^2 Q_1 Q_5 \cos^2\theta}{h^2 f^2}\right) \cos^2\theta d\tpsi^2+\left(r^2+a^2 -\frac{a^2 Q_1 Q_5 \sin^2\theta}{h^2 f^2} \right) \sin^2\theta d\tphi^2\right]\,,
\end{eqnarray}
with
\begin{equation}
 f=r^2+a^2\cos^2\theta\,, \quad h=\sqrt{H_1 H_5}=\left[\left(1+\frac{Q_1}{f}\right) \left(1+\frac{Q_5}{f}\right)\right]^{1/2}\,.
\end{equation}
In the near-horizon limit, $r \ll (Q_1 Q_5)^{1/4}$, $a\ll (Q_1 Q_5)^{1/4} \ll R$, this solution is dual to a Ramond ground state with maximal R charge. To see this, we remind the reader that spectral flow maps the Ramond ground state to the Neveu-Schwarz ground state and that this flow is implemented by the large coordinate transformation
 \begin{equation}\label{eq:geomSF}
 \psi=\tpsi -\frac{y}{R}\,, \quad \phi=\tphi -\frac{t}{R}\,.
\end{equation}
Applying \eqref{eq:geomSF} to the above metric yields
\begin{equation}\label{eq:metricNS}
 ds^2_{NS}=\sqrt{Q_1 Q_5}\left[-(r'^2+1)\frac{dt^2}{R^2}+r'^2\frac{dy^2}{R^2}+\frac{dr'^2}{r'^2+1}+d\theta^2+\cos^2\theta d\psi^2+\sin^2\theta d\phi^2\right]\,.
\end{equation}
This is just global $AdS_3 \times S^3$ and is indeed dual to the NS vacuum state as desired.

\subsubsection{Evanescent ergosurface}

The ERS instability relies on a key feature of supersymmetric microstate geometries dubbed the evanescent ergorsurface in \cite{Gibbons:2013tqa}. To describe this surface, recall \cite{Witten1981} that
supersymmetry implies the existence of a globally null Killing vector field which, when there exists a Kaluza-Klein Killing field $\partial_y$, may be written
\begin{equation}\label{eq:Killing}
 V=\partial_t+\partial_y\,.
\end{equation}
Here $\partial_t$ and $\partial_y$ are commuting Killing vector fields. The Killing field $\partial_y$ is spacelike and is associated with the Kaluza-Klein direction of the 6d geometry, while $\partial_t$ becomes timelike and canonically normalized near infinity.  As a result, $V$ can also be related to a non-spacelike Killing vector of the 5d geometry obtained from dimensional reduction along the $y$ circle.  Since $V$ is globally null it is everywhere tangent to affinely parametrized null geodesics.  It will be convenient to refer to $V$ as the SUSY Killing field below.

The evanescent ergosurface $\mathcal{S}$ is then defined by $V \cdot \partial_y=0$.  It is thus located at $f=0$ in the geometry \eqref{eq:metricR}, where $r=0$ and $\theta=\pi/2$. Hence $\mathcal{S}$ is a 2d timelike submanifold of the 6d geometry. At this locus the Kaluza-Klein circle $y$ pinches off smoothly, as does $\psi$. At constant $t$ the topology of $\mathcal{S}$ is $S^1$ where the coordinate around this circle is $\phi$. The Killing vector field $\partial_t$ is timelike everywhere except on $\mathcal{S}$ where it is null ($V$ is null everywhere and $\partial_y$ vanishes on $\mathcal{S}$). There are zero-energy null geodesics with tangent vector $V$ which are stably trapped on $\mathcal{S}$ and thus stay at constant $(r,\theta)=(0,\pi/2)$; more on this in~\S\ref{sec:instability}. This evanescent ergorsurface will be the location of our Aichelburg-Sexl pp-wave.

\section{Adiabatic instability of 2-charge microstate geometries}\label{sec:instability}

We are now ready to add null particles moving in the $\phi$ direction of the $S^3$ at $\theta=\pi/2$ and at the center of $AdS_3$ ($r=0$).  This is the location of the evanescent ergosurface after spectral flow.  Our focus will be on studying the backreaction induced by such particles.  

From the CFT perspective, the addition of a particle corresponds to exciting higher harmonics $N_k$.
Starting with the NS vacuum or, after spectral flow, the Ramond ground state with maximal R-charge, we will see in~\S\ref{sec:instability} that the instability found in~\cite{Eperon:2016cdd} will take us towards more complex and typical states $\ket{\{N_k\}}$. Our main focus, however, is on explaining the physical implications of the instability found in~\cite{Eperon:2016cdd} for the gravity solutions \eqref{eq:fuzzgeos}.  We therefore begin with a brief review of this instability.

\subsection{The ERS instability}\label{ssec:ERS}

The instability identified in \cite{Eperon:2016cdd} is a consequence of a property called stable trapping, which is exhibited by the microstate geometries near the evanescent ergosurface $\mathcal{S}$ where the SUSY Killing field $V$ is tangent to affinely parameterized null geodesics with zero energy. These geodesics are at rest relative to infinity, in contrast to the microstate geometries which have a non-zero angular momentum. This implies that particles following orbits of $V$ resist the frame-dragging effect caused by the rotation of the background geometry. In this sense, the zero-energy null geodesics can be seen as possessing angular momentum opposite to that of the microstate geometry. These geodesics remain within the bounded region of the evanescent ergosurface and are thus trapped. Because they sit at the bottom of a gravitational potential well they minimize the energy and so the trapping phenomenon is stable.

Now imagine perturbing the spacetime by adding an uncharged massive particle near to the evanescent ergosurface. If we neglect backreation, the particle moves on a geodesic. When coupled to supergravity fields it will gradually radiate energy and angular momentum and its trajectory will approach a geodesic that minimizes the energy. Hence the trajectory of the particle will approach one of the zero-energy trapped null geodesics tangent to $V$ on the evanescent ergosurface. The particle will have very small energy as measured at infinity but, since the massive particle is now following an almost null trajectory, the energy measured by a local observer will be very large.  It will thus give rise to strong backreaction. As argued in~\cite{Eperon:2016cdd}, this suggests an instability that triggers a large change in the spacetime geometry.

While the above reasoning used particles, one should obtain the same conclusions using a field-theoretic analysis in the WKB limit, and analogous physics follows from studying quasi-normal modes \cite{Eperon:2016cdd}.  In the particle context, the fact that interactions played an important role (by allowing the massive particle to radiate) means that the instability is a non-linear effect.  Note that the instability is fundamentally a consequence of the existence of stably trapped null geodesics and that an evanescent ergosurface per se is not required.  In particular, one expects this instability to arise even in supersymmetric microstate geometries that do not possess a Kaluza-Klein Killing vector field and thus no concept of an evanescent ergosurface.  In this sense, the ERS instability appears to be a rather robust feature of supersymmetric microstate geometries.

What could be the endpoint of this instability? Its overall effect is to remove angular momentum from the microstate geometry via radiation. This will cause the evanescent ergosurface to shrink. It was suggested in~\cite{Eperon:2016cdd} that a natural endpoint is a non-supersymmetric black hole with the same conserved charges as the microstate geometry but different angular momenta.

We will now argue for a different conclusion.  To do so, we recall \cite{Eperon:2016cdd} that orbits of the SUSY Killing field $V$ on the evanescent ergosurface are null geodesics.  We then return to the above discussion of adding a particle and consider the limit where the particle becomes massless and travels precisely along such a geodesic.  Such particles preserve the supersymmetry of the background geometry, so in this limit one expects there to be a stationary supergravity solution that incorporates the full backreaction from the particle even when the local energy and momentum of the null particle are large.  This is not to say that the ERS instability has been completely removed, as even tiny deformations away from this limit will still trigger its effects.  However, continuity implies that the ERS instability proceeds very slowly when the system is close to this SUSY null particle limit.  Furthermore, we recall that the ERS instability tends only to make the particle more null and to move it even closer to the above null geodesics while increasing the locally-measured energy.  As a result, close to our SUSY null limit, one may approximate the evolution induced by the ERS instability as adiabatic evolution along a one-parameter family of fully-backreacted supersymmetric supergravity solutions describing null particles on the above SUSY geodesics.  The natural parameter labeling the solutions is just the locally-measured energy of the null particle, and dynamical evolution drives this energy to slowly increase.

Our first task is thus to identify the relevant supergravity solutions.  As is well known, the backreaction of a null particle in flat space is described by the Aichelburg-Sexl solution \cite{Aichelburg1971}, which preserves the desired supersymmeries \cite{Berenstein2002}.  We therefore seek supersymmetric solutions of the D1-D5 system which locally take the Aichelburg-Sexl form near the null geodesic on which the particle travels.  To simplify the analysis, we will in fact consider a more symmetric situation describing an ensemble of such particles that preserves both translation invariance on the internal $T^4$ and rotational invariance under $\partial_\phi$: in the language commonly used to describe such solutions, we smear the particles over these directions. It will be convenient to begin with the maximally rotating microstate and in fact to start our discussion in the near-horizon limit which, under the spectral flow transformation discussed in \S~\ref{smoothgeom} becomes just $AdS_3\times S^3$.

\subsection{Aichelburg-Sexl solutions}\label{ssec:ASinAdS}

We therefore consider the addition  to $AdS_3\times S^3$ of an Aichelburg-Sexl shock wave associated with a ring of particles moving at the speed of light around a circle on the $S^3$ at the center of AdS.  As shown in \cite{Lunin:2002fw,Lunin:2002bj}, the resulting geometry is

\begin{eqnarray}\label{eq:ASmetricNS}
 d\bar{s}_{NS}^2&=&\sqrt{Q_1 Q_5}\left[-(r'^2+1)\frac{dt^2}{R^2} + r'^2 \frac{dy^2}{R^2}+\frac{dr'^2}{r'^2+1} + d\theta^2+\cos^2\theta d\psi^2+\sin^2\theta d\phi^2 \right]\nn\\
 &&+\frac{q \sqrt{Q_1 Q_5}}{r'^2+\cos^2 \theta} \left[\left((r'^2+1) \frac{dt}{R}+\sin^2\theta d\phi\right)^2-\left(r'^2 \frac{dy}{R}-\cos^2\theta d\psi\right)^2\right]\,,
\end{eqnarray}
where we have corrected some typos in the expressions of \cite{Lunin:2002fw,Lunin:2002bj}.
 In \eqref{eq:ASmetricNS}, $q$ parametrizes the locally-measured energy of the null particle; i.e., it describes the strength of the shock.  For $q=0$ \eqref{eq:ASmetricNS} is empty $AdS_3 \times S^3$ as desired.

The geometry~\eqref{eq:ASmetricNS} has a curvature singularity at the locus of the shockwave.
Near the evanescent ergosurface $(r,\theta)=(0,\pi/2)$, the leading terms in~\eqref{eq:ASmetricNS} yield
\begin{equation}
\label{eq:ASzoom}
 d\bar{s}_{NS}^2=\sqrt{Q_1 Q_5} \left[-\frac{dt^2}{R^2}+dr'^2+d\theta^2+d\phi^2+\frac{q}{f}\left(\frac{dt}{R}+d\phi\right)^2\right]\, ,
\end{equation}
which is precisely an Aichelburg-Sexl shock in otherwise-flat space propagating along $\tphi=\phi+\frac{t}{R}$.  Note that, as for the 2-charge geometry without the shockwave~\eqref{eq:metricR}, the $y$ and $\psi$ circles pinch off at $f=0$.

It is now straightforward to invert the spectral flow~\eqref{eq:geomSF} and obtain the R sector solution.  We further restore the asymptotically flat region by judiciously adding back the appropriate constants inside the harmonic functions. Defining the parameter $\xi = 1-q$, this construction suggests that taking the maximally-rotating geometry~\eqref{eq:metricR}, adding a ring of particles to the evanescent ergosurface and incorporating their backreaction, one obtains the geometry
\begin{eqnarray}\label{eq:ASmetricR2}
d\bar{s}^2_R&=&-\frac{1}{\bar{h}} (dt^2-dy^2) + \bar{h}\bar{f} \left(d\theta^2+\frac{d\bar{r}^2}{\bar{r}^2+\bar{a}^2} \right) -\xi\frac{2a \sqrt{Q_1 Q_5}}{\bar{h}\bar{f}} \left(\cos^2\theta dy d\tpsi + \sin^2\theta dt d\tphi \right)\\
 &&+\bar{h}\left[\left(\bar{r}^2+\xi \frac{\bar{a}^2 Q_1 Q_5 \cos^2\theta}{\bar{h}^2 \bar{f}^2}\right) \cos^2\theta d\tpsi^2+\left(\bar{r}^2+\bar{a}^2 -\xi \frac{\bar{a}^2 Q_1 Q_5 \sin^2\theta}{\bar{h}^2 \bar{f}^2} \right) \sin^2\theta d\tphi^2\right]\,,\nn
\end{eqnarray}
where
\begin{equation}
  \bar{h}=\sqrt{\bar{H}_1\bar{H}_5}=\left[\left(1+\frac{Q_1}{\bar{f}}\right) \left(1+\frac{Q_5}{\bar{f}}\right)\right]^{1/2}\,, \quad \bar{f}=\bar{r}^2+\bar{a}^2\cos^2\theta=\xi f\,.
\end{equation}
One can show that \eqref{eq:ASmetricR2} is generated by the string profile
\begin{eqnarray}
\label{ASstring}
 \vec{\bar{F}}(v)&=&(\bar{a} \cos(\omega v/\xi+\phi_0),\bar{a} \sin(\omega v/\xi+\phi_0),0,0)\,, \quad 0\leq v\leq L\xi\nonumber\\
 \vec{\bar{F}}(v)&=&(\bar{a}\cos\phi_0,\bar{a}\sin\phi_0,0,0)\,, \quad L \xi \leq v < L\,
\end{eqnarray}
after smearing over $\phi_0$ \cite{Lunin:2002bj}. The smearing operation should be understood as generalizing \eqref{eq:fuzzgeos} by adding further terms to the harmonic functions sourced by a set of independent string profiles $\vec{F}_i$ with independent values of $\phi_0$ and then taking a limit where the profiles in fact coincide and the ensemble of $\phi_0$ values forms the uniform distribution on $[0,2\pi]$.  This construction makes it clear that the result \eqref{ASstring} is indeed an appropriately supersymmetric solution, once augmented by the appropriate dilaton and form fields generated by \eqref{ASstring}.\footnote{While the profile function \eqref{ASstring} is very similar to the profile function that generates the solutions dual to spectral flows of the conical deficits  \cite{Balasubramanian:2000rt,Maldacena:2000dr,Balasubramanian:2005qu}, it has a different range of integration which destroys the Hopf structure that leads to the conical singularity.  With \eqref{ASstring} one finds a curvature singularity instead.} Readers concerned about the breakdown of the supergravity description near the shock may think of \eqref{ASstring} as an approximation to a smooth profile whose Fourier decomposition has no excitations higher than the $N^\text{th}$ harmonic.

Returning to the string profile \eqref{ASstring},
before smearing one sees that the profile describes a string that winds once around the $\phi$-circle on the interval $v\in[0,L\xi]$ and then remains at the same $x$-location for the remaining $v$-length $(1-\xi)L$. The last straight segment corresponds to the added particle: just a bump on a fuzzball.~\footnote{We are grateful to Iosif Bena for emphasizing this viewpoint.}
From this profile one obtains the harmonic functions~\cite{Lunin:2002bj}
\begin{eqnarray}
 \bar{H}_5&=&1+\frac{{Q}_5\xi}{\bar{r}^2+\bar{a}^2 \cos^2\theta}+\frac{{Q}_5(1-\xi)}{(x_1-\bar{a}\cos\phi_0)^2+(x_2-\bar{a}\sin\phi_0)^2+x_3^2+x_4^2}\,,\\
 \bar{H}_1&=&1+\frac{Q_5 \bar{a}^2 \omega^2 /\xi}{\bar{r}^2+\bar{a}^2 \cos^2\theta}\,,\\
 \bar{A}_{\tphi}&=& -Q_5 \bar{a}^2 \omega\frac{\sin^2\theta}{\bar{r}^2+\bar{a}^2 \cos^2\theta}\,,
\end{eqnarray}
where the radial coordinate at infinity $\bar{r}$ is related to $r$ by
 \begin{equation}
  \bar{r}=\sqrt{\xi}\,r\, ,
\end{equation}
so that $\epsilon_{\bar{r}\theta \tphi \tpsi}=\sqrt{g}=(\bar{r}^2+\bar{a}^2\cos^2) \bar{r} \sin\theta \cos\theta$ and the flat metric takes the form
\begin{equation}
 dx_4^2 = (\bar{r}^2+\bar{a}^2\cos^2\theta)\left(\frac{d\bar{r}^2}{\bar{r}^2+\bar{a}^2} + d\theta^2\right) + (\bar{r}^2+\bar{a}^2) \sin^2\theta d\tphi^2 + \bar{r}^2 \cos^2 \theta d\widetilde  \psi^2\,.
\end{equation}
Averaging over $\phi_0$ gives
 \begin{equation}\label{eq:ASharmonics}
  \bar{H}_5=1+\frac{{Q}_5}{\bar{r}^2+\bar{a}^2 \cos^2\theta}\,, \quad \bar{H}_1=1+\frac{{Q}_1}{\bar{r}^2+\bar{a}^2 \cos^2\theta}\,, \quad \bar{A}_{\tphi}=-{Q}_5 \bar{a}^2 \omega \frac{\sin^2\theta}{\bar{r}^2+\bar{a}^2 \cos^2\theta}\,,
 \end{equation}
which leads to the geometry~\eqref{eq:ASmetricR2}.
Note that the relation~\eqref{eq:Q1} yields
 \begin{equation}
\label{eq:key}
  \bar{a}=\sqrt{\xi}\, a\,.
 \end{equation}
Though it seems innocent enough, this equation is actually key to our analysis.  It implies the backreacted solution to be scaled down by a factor $\sqrt{\xi}$.

\subsection{The shrinking shockwave}\label{ssec:shrinkingAS}

We argued above that the ERS instability admits an adiabatic limit described by the family of solutions \eqref{eq:ASmetricR2} with increasing strength $q$ of the Aichelburg-Sexl shock, and thus with decreasing $\xi$.   From~\eqref{eq:J} and the asymptotics of the metric~\eqref{eq:ASmetricR2} one finds that the angular momentum of any such solution is smaller than in the maximally rotating case by a factor $\xi=1-q\leq 1$ while the total energy is unchanged. We find
\begin{equation}
 \bar{\mathcal{E}}=Q_1+Q_5\,, \qquad \bar{J}_{\tphi}=\xi Q_1 Q_5\,
\end{equation}
which corresponds to $j=\xi j_{\rm max}=\xi n_1 n_5$.
Since \eqref{eq:ASmetricR2} still possesses an evanescent ergosurface, the solution will continue to shrink and radiate angular momentum to infinity so long as the ERS analysis remains valid.  Indeed, while a consistent supergravity description will require a series of duality frames as we decrease $j$ \cite{Chen:2014loa},
the existence of a (perhaps singular) evanescent ergosurface is guaranteed in all frames by the supersymmetry of the solution.

The solution will continue to shrink at least until we can no longer trust the ERS analysis at $\xi\sim 1/\sqrt{n_1n_5} = 1/\sqrt{N}$. In the large $N$ limit this corresponds to taking $\xi\rightarrow 0$, which gives
\begin{equation}
 \bar{a}=\sqrt{\xi} a \to 0\,, \quad \bar{f}=\bar{r}^2+\xi a^2 \cos^2\theta \to \bar{r}^2\,, \quad \bar{h}=\left[\left(1+\frac{Q_1}{\bar{f}}\right)\left(1+\frac{Q_5}{\bar{f}}\right)\right]^{1/2} \to \frac{\sqrt{Q_1 Q_5}}{\bar{r}^2}\,.
\end{equation}
In this limit we recover the near-horizon metric of the $M=0$ extremal BTZ black hole with transverse $S^3$ \cite{Banados:1992gq,Banados:1992wn}:
\begin{equation}
\label{eq:naivegeo}
 d\bar{s}_{R}^2=\frac{\bar{r}^2}{\sqrt{Q_1 Q_5}} (-dt^2+dy^2)+\sqrt{Q_1 Q_5}\left(\frac{d\bar{r}^2}{\bar{r}^2}+d\theta^2+\cos^2\theta d\tpsi^2+\sin^2\theta d\tphi^2\right)\,.
\end{equation}
This is consistent with the ERS suggestion that the system evolves to become a black hole.

One effect not taken into account by ERS is the possibility that the particle seeding the instability will decay.  So long as the decay products continue to be treated as classical particles, one presumes this to give rise to a set of ERS-like instabilities all acting in concert.  But since we consider a limit where the instability is adiabatically slow, this system of particles will reach some sort of equilibrium at each $j$.  Indeed, in the absence of other constraints, a coarse-grained description of this equilibrium should resemble the microcanonical ensemble of all appropriately supersymmetric states with the given value of $j$; after including backreaction, this is just the microcanonical ensemble of microstate geometries.

The ERS analysis thus suggests that there is a general tendency for asymptotically flat microstate geometries to evolve towards smaller $j$.  This is no surprise for $j \sim j_\text{max}$, as the microcanoncal entropy $S(j)$ decreases with increasing $j$ in this regime according to \eqref{eq:Slargej}. In fact, $S(j)$ behaves this way for all $j > j_\text{typical}$, and so {\it any} interaction should lead to this behavior when the microstate is well-described by supergravity.

On the other hand, we recall from \S~\ref{sec:cft} that $S(j)$ is maximized at $j_\text{typical}$ of order $\sqrt{n_1n_5}$.  As a result, so long as our microcanonical ensemble approximation remains valid and the entropy in radiation at infinity can be neglected\footnote{This is a subtle point.  The entropy of radiation at infinity is divergent.  We may regulate the model by placing the system in a finite-sized box. Then near $j_\text{typical}$, in the limit of large charges $n_1,n_5$ with fixed box size, the entropy in the radiation is negligible when compared with the microstate density of states $S(j$).},  unitarity prohibits {\it any} interaction from causing $j$ to decrease below $j_\text{typical}$.  This strongly suggests that -- at least for generic microstates -- the ERS mechanism shuts down for $j$ near $j_\text{typical}$. The effect of the ERS instability is thus to drive smooth solutions towards stringy typicality - a rough end for these supposedly smooth spacetimes.

There is indeed ample room for corrections to the ERS analysis in this regime.  As noted in \S~\ref{sec:geometry}, microstate geometries with $j\sim j_\text{typical}$ have string-scale structure and could well require large corrections to the classical supergravity description used by ERS.  While a full analysis is beyond the scope of this work, we describe a particular stringy effect in \S~\ref{ssec:qmmodel} below that could plausibly provide such corrections and illustrate the resulting stabilization in a simple toy model.

\section{A model for stabilization at typicality}\label{ssec:qmmodel}

The ERS analysis considered test particles and fields propagating on microstate geometries. At large $j$ the geometries are quite smooth, so stringy corrections can be incorporated via an asymptotic expansion in $\alpha'$. However, due to the presence of string-scale structure when $j\sim j_\text{typical}$, an accurate analysis in this regime requires any probes to be treated as quantum strings.  In particular, the zero-point oscillations of probe strings mean that they will not sit sharply at the minimum of any background potential.  One may thus expect this effect to raise the energy of the probe above what would be expected by naively extrapolating results from the smoother geometries at larger $j$.  As a result, this mechanism has the potential to deactivate the ERS instability at $j \sim j_\text{typical}$.  While a complete stringy analysis is beyond the scope of our work, we provide a simple toy model below exhibiting what we believe to be key features of the physics.

To set the context for our model, let us briefly return to the ERS discussion of massive particles.  As discussed in \S~2.4 of ERS, the energy of such particles is minimized at $j_{particle} = -\infty$ in the geometry \eqref{eq:metricR} with $j=j_\text{max}$.  In particular, the minimum of the energy $E_\text{min}(j_{particle})$ decreases as $j_{particle} \rightarrow -\infty$ and so the particle tends to roll down this effective-potential hill by radiating into the asymptotically flat region.

Of course, once $j_{particle}$ becomes large one must take backreaction into account.  One would then like to compute the minimum energy $E_\text{backreacted, min}(j)$ consistent with a given total angular momentum $j$ (including $j_{particle}$) and the existence of the particle.  Doing so will be complicated away from the adiabatic limit of \S~\ref{sec:instability}, but one expects the result  to give an effective potential $E_\text{backreacted, min}(j)$ whose qualitative features are similar to the above $E_\text{min}(j_{particle})$, and in particular which again decreases as we make $j$ more negative.

A toy model for such an effective potential computation is given by a family of one-dimensional models in non-relativistic quantum mechanics defined by potentials $V_j(x)$ for which we wish to compute the energy $E_\text{model, min}(j)$ of the ground state.  We consider the Hamiltonian
\begin{equation}
H = \frac{p^2}{2m} + V_j(x)
\end{equation}
for each value of a parameter that we will also call $j$.  Here there is no explicit notion of backreaction, though it has been incorporated implicitly through our comparison of ground state energies for different values of the external parameter $j$.

\begin{figure}[h!]
\centering
{\includegraphics[width=.5\textwidth]{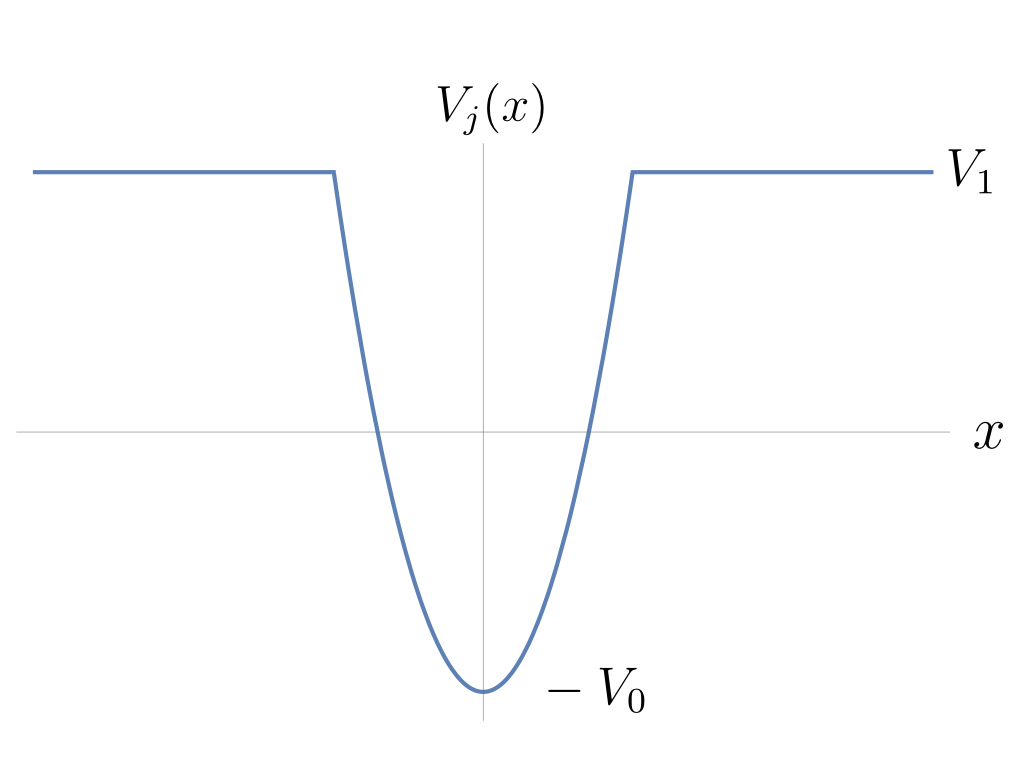}}
\caption{The potential in our toy model. The real shape of the effective potential for timelike particles in a microstate geometry is a smoothed version.
}
\label{Vs}
\end{figure}

One would like this potential to model the effective potential for timelike particles in a microstate geometry, which is minimized at the evanescent ergosurface and which becomes constant far away.  It thus takes the general shape of the potential in figure \ref{Vs}.  For simplicity, we model this shape by choosing

\be
\label{1particleV}
V_j(x)= \left\{
	\begin{array}{ll}
		\frac{1}{2} m\omega(j)^2 x^2 - V_0(j)  &\quad \quad |x|<L \\
		V_1(j) & \quad \quad |x|>L
	\end{array}
	\right. .
\ee
$L$ characterizes the scale over which the potential differs from its asymptotic value, and continuity of the potential requires
\be
\frac{1}{2} m\omega^2 L^2 - V_0 = V_1.
\ee
To model the ERS instability, all the parameters should depend on $j$ except the particle mass $m$. We will often leave this functional dependence implicit.

Near $x=0$ the eigenfunctions match the harmonic oscillator, but the effects of the flat potential in the $|x|>L$ region begin to affect the $n^\text{th}$ and higher states when the position fluctuations
\begin{equation}
\langle x^2 \rangle\approx \frac{(2n+1)}{2m \omega}
\end{equation}
become $O(L^2)$. In particular, in states with $\langle x^2\rangle \gg L^2$ the particle will not be bound to the harmonic trap. It will be useful to define the dimensionless quantity
\begin{equation}
\label{defC}
C := m\omega L^2
\end{equation}
which essentially counts the number of bound states in the potential $V_j(x)$: states with excitation number $n \ll C$ are well-approximated by harmonic oscillator eigenstates, while those with $n \sim C$ are still bound but receive significant corrections from the turning point in the potential \eqref{1particleV}. For $n \gg C$ the particle is effectively free.

As $j$ decreases, the length scales of structures in our potential should decrease in analogy with the decreasing size of structures in the microstate geometries. Thus we take $\omega$ to increase, and in order to keep the number of bound states constant we hold $\omega L^2$ fixed (this is the natural scaling in non-relativistic quantum mechanics).  We take $V_0$ to slowly decrease with $j$ in order to make the ground state energy $E_\text{model, min}(j)$ behave like $E_\text{min}(j_{particle})$, and in particular to slowly drive the solution towards smaller $j$.

So far we have merely constructed a simple quantum-mechanical toy model of the original ERS particle analysis. However, we wish to consider effects associated with the zero-point oscillations of stringy probes of the microstate geometries. In our model this can be accomodated by letting the test particle have internal structure.  For the present purposes, it will be enough to regard the particle as a bound state of $K$ partons (say, each of mass $m/K$) coupled by an additional internal potential $\epsilon_\text{int}$ that depends only on the relative separations of the partons and not on $j$. If one likes, one may take these $K$ particles to be connected by springs in a ring in order to give a discrete model of a quantum string.

Since the potential $V_j$ largely models gravitational redshift effects in each microstate background, we will take each parton to experience the same potential $V^{parton}$ whose parameters $\omega_{parton}, L_{parton}, V_0^{parton}$ we fix below in terms of the parameters of the particle model \eqref{1particleV}.  The full Hamiltonian is
\be
\label{eq:qm1p}
H = \sum_{i=1}^{K} \left(\frac{Kp_i^2}{2m} + V^{parton}(x_i)\right) + \epsilon_\text{int}.
\ee
We begin in the regime where the external potential $\omega_{parton}$ is small compared to all scales in the internal potential $\epsilon_\text{int}$.  This models microstates, like the maximally-rotating solution, whose structures are large compared to the string scale.  In this regime the internal degrees of freedom are effectively in their ground state and we obtain a ``tight binding'' limit in which any differences between the $x_i$ are small compared to any scales in the external potential.  The result is an effective description of the parton composite as a single particle of mass $m$ moving in a 1-particle potential $KV^{parton}$ evaluated at the center of mass coordinate $\bar{x}$. The effective physics exactly matches the single-particle model above if we identify
\be
V^{parton} =\frac{V_j}{K}.
\ee
This implies $L_{parton} = L$, $\omega_{parton} = \omega$ and $KV_0^{parton} = V_0$.

So long as $C>1$, there is a ground state bound to the well in which
\be
\langle\bar{x}^2\rangle \approx \frac{1}{2m\omega}.
\ee
The harmonic oscillator approximation to $V(\bar x)$ implies that the ground state energy of the composite system is
\be
E_0 \approx E_{\text{tight binding}} := \frac{\omega}{2} - V_0 + \epsilon_0,
\ee
where $\epsilon_0$ is the ground state energy of Hamiltonian describing the intra-parton couplings.

However, the properties of the model become very different at $\omega\gg \epsilon_\text{int}$, i.e. as $j$ decreases towards typicality. Any bound partons are much more strongly coupled to the external potential than to each other; if the partons remained bound, the ground state of the composite system would have each parton separately in the ground state of the potential $V^{parton}$.   However, defining $C_{parton}$ in analogy with \eqref{defC} yields
\begin{equation}
\label{compareCs}
C_{parton} : = \frac{m}{K} \omega_{parton} L_{parton}^2 = \frac{C}{K}.
\end{equation}
This is the quantity that counts states bound to the external potential when interactions between partons can be ignored.  Taking $K\gtrsim C$ partons, the number of such bound states will become less than one in this regime and it will be inconsistent to continue to treat all partons as bound in the external potential.

Instead, the partons pop out of the external potential well and
experience only the flat potential $V_1^{parton} = V_1/K$ to good approximation when $K\gg 1$.\footnote{At any given time, some of the partons will in fact lie within their potential well.  This effect can be estimated by studying the effective potential $K\langle  V^{parton}\rangle_{\bar x}$, where the notation indicates the expectation value of $V^{parton}$ for some one parton in the approximation that $\bar x$ is held fixed but that the system is otherwise in its ground state.  One finds it to be of order $1/\sqrt{K}$, so we neglect it.}  As a result, the actual ground state energy in this regime
will be
\be
\label{Eout}
E_{\text{model, min}} \approx V_1 + \epsilon_\text{int}   = \frac{1}{2} m \omega^2 L^2 - V_0+ \epsilon_\text{int} =
E_{\text{tight binding}} +(\epsilon_\text{int}-\epsilon_0)+ \frac{\omega}{2}(C - 1).
\ee
Taking $C > 1$ so that there is at least initially a bound state, the corrections to the tight binding energy are positive. They scale with $\omega$ at large $\omega$ and so counteract any tendency of $E_{\text{tight binding}}$ to slowly decrease due to the $j$-dependence of $V_0$. The behavior at smaller $K$ is similar.

Note that the analogue of the ERS effective potential is $E_{\text{tight binding}}$, and that this generally differs from the actual ground state energy that would arise from putting all the
particles inside the external potential well.  The latter knows
about the internal structure of the composite particle, while the ERS potential does
not.  Writing $E_{\text{model, min}}$ in terms of $E_{\text{tight binding}}$ clearly displays the extra positive term that exhibits stabilization.

To summarize, in our toy model decreasing $j$ causes the ground state energy to decreases for a while as the instability proceeds.  However, it then begins to increase again when the zero-point oscillations of the probe string no longer fit into the external potential well.  Analogous behavior for the ERS phenomenon would mean that the instability stabilizes when the evanescent ergosurface develops string-scale structure, which occurs as the CFT state approaches typicality.

\section{Discussion}\label{sec:conclusions}

We have argued that an adiabatic limit of the ERS instability of the 2-charge D1-D5 system is described by motion along a family of microstate geometries associated with the D1-D5 CFT.  In particular, due to the emission of radiation to infinity, the angular momentum labelling the relevant microstate geometries should be thought of as a slowly-evolving function of time $j(t)$. When the instability is very weak and this evolution is especially slow, there is time for any perturbation to induce transitions between microstates and the geometry at any time $t$ should admit an approximate description as the ensemble of all supersymmetric geometries with angular momentum $j(t)$, described in \cite{Balasubramanian:2008da}.    At large $j$ the ERS instability is consistent with entropic reasoning in the CFT and indeed could have been anticipated on such grounds. From the field theory point of view, the instability simply causes evolution from states described by rare collections of twist operators to those described by more generic such collections.

On the other hand, entropic reasoning suggests that the instability terminates when $j$ approaches $ j_\text{typical} \sim \sqrt{n_1n_5}$. Since this is also the regime where stringy corrections to \cite{Eperon:2016cdd} naturally become large, we suggested that the system is indeed stabilized at such $j$.   A plausible scenario is that the zero-point oscillations of any perturbing string then prohibit it from taking full advantage of the strong redshift near the evanescent ergosurface as this surface also exhibits string-scale structure.  A full analysis is beyond our scope, but the toy model of \S~\ref{ssec:qmmodel} illustrates how this effect might tame the instability.

It is important to emphasize that we have argued for stabilization only in our adiabatic limit.  Since the ERS instability is non-linear, it will evolve quickly under large perturbations that take the system far away from the supersymmetric moduli space.  It appears difficult to analyze this regime, and one could well imagine the endpoint in the case being either a horizon-free (but not smooth) solution with string-scale structure (a.k.a. a {\it rough} microstate), or a traditional black hole.  As usual in this field, the question remains open for future investigation.

It would be interesting to consider a similar analysis for the 3-charge system.  While in that setting it is unclear that there is any geometric analogue of typical microstates, one may in any case choose to study known classes of geometric solutions.  Some initial steps involving the addition of Aichelburg-Sexl shockwaves to one such family are taken in appendix \ref{app:3charge}, but it remains to check that the conjectured fields do in fact satisfy the supergravity equations of motion, or to study more typical 3-charge microstates \cite{Bena:2016ypk}.

Even with our presumed stabilization at $j \sim j_\text{typical}$,  the fact that it modifies the ERS instability only when the supergravity description breaks down means that much of the physical interpretation of ERS remains intact:  the slightest perturbation will cause microstates with large angular momentum to collapse, with the likely endpoint being (geometrically) indistinguishable from the $M=0$ BTZ black hole. This does not prevent one from preparing the black hole in such a microstate but, depending on parameters, it could well cause the microstate to collapse and absorb the observer into its structure before she can sail through any smooth region where the spacetime caps off.

\section*{Acknowledgements}
We would like to thank Iosif Bena, Roberto Emparan, Henry Maxfield, Harvey Reall, Jorge Santos, and David Turton for useful discussions.
DM was supported in part by the U.S. National Science Foundation under grant number PHY15-04541 and by funds from the University of California. BM was supported by NSF Grant PHY13-16748.
AP acknowledges support from the Black Hole Initiative (BHI) at Harvard University, which is funded by a grant from the John Templeton Foundation and support from the Simons Investigator Award 291811 and the DOE grant DE-SC0007870.

\appendix

\section{Instability of 3-charge microstate geometries}\label{app:3charge}

We now discuss the ERS instability for the special class of 3-charge geometries constructed in~\cite{Lunin:2004uu,Giusto:2004id,Giusto:2004kj,Giusto:2004ip} and studied also by ERS \cite{Eperon:2016cdd}. Building on the 2-charge solution of~\S\ref{sec:2charge} (but now with rotation along both angles $\phi$ and $\psi$ of the $S^3$ turned on) dual to Ramond ground states, the action of spectral flow~\eqref{eq:CFTSF} with $\alpha \neq \pm 1$ yields excited states. In addition to D1 and D5 brane charge, these solutions have momentum excitations along the common D1-D5 direction. We review this special class of 3-charge solutions from the CFT and geometry descriptions and then briefly discuss the ERS instability along the same lines as~\S~\ref{sec:instability}.

\subsection{CFT}
States in the D1-D5 CFT with momentum excitations along the common $y$ direction correspond to excited Ramond sector states. Starting with the Neveu-Schwarz vacuum we can generate excited states in the Ramond sector through the action of spectral flow~\eqref{eq:CFTSF}.
The 3-charge states of interest are obtained by acting on the Neveu-Schwarz vacuum in the left-moving sector with
\begin{equation}\label{eq:CFTSFalpha}
 \alpha=2n+1 \quad \text{with} \; n \; \text{integer}\,,
\end{equation}
and in the right-moving sector with $\alpha=1$ (so that the right movers are in their Ramond ground state and the CFT is supersymmetric).
After spectral flow~\eqref{eq:CFTSF} with~\eqref{eq:CFTSFalpha} the states in the symmetric product theory have dimensions $(h,\tilde{h})$ and charges $(j,j')$:
\begin{eqnarray}
 &h=\frac{1}{4}(2n+1)^2 n_1 n_5\,,\quad \tilde{h}=\frac{1}{4} n_1 n_5\,,&\\
& j=-\frac{1}{2}(2n+1) n_1 n_5\,, \quad j'=-\frac{1}{2} n_1 n_5\,.&
\end{eqnarray}
We get D1-D5-$p$ states carrying momentum charge
\begin{equation}\label{eq:np}
 n_p=h-\tilde{h}=n(n+1)n_1 n_5\,,
\end{equation}
along the $S^1$
and angular momenta
\begin{equation}
 j_\psi=-j'+j=-n \,n_1 n_5\,, \quad j_\phi=-j'-j=(n+1)n_1 n_5\,,
\end{equation}
on the angles of the $S^3$.

\subsection{Geometry}

The special class of 3-charge solutions obtained from the spectral flow~\eqref{eq:geom3SF} of the maximally rotating 2-charge solution~\eqref{eq:metricR} are given by \cite{Giusto:2004id,Giusto:2004ip}
\begin{eqnarray}\label{eq:3metricR}
 ds^2_R&=&-\frac{1}{h}(dt^2-dy^2) +\frac{Q_p}{h f} (dt-dy)^2 +hf \left(\frac{dr^2}{r^2+(\tg_1+\tg_2)^2 \eta}+d\theta^2\right)\nn\\
 &&+h\left(r^2+\tg_1(\tg_1+\tg_2)\eta -\frac{(\tg_1^2-\tg_2^2)\eta Q_1 Q_5 \cos^2\theta}{h^2 f^2} \right) \cos^2\theta d\tpsi^2\nn\\
 &&+h\left(r^2 +\tg_2(\tg_1+\tg_2)\eta+ \frac{(\tg_1^2-\tg_2^2)\eta Q_1 Q_5 \sin^2\theta }{h^2f^2} \right) \sin^2\theta d\tphi^2\nn\\
 &&+ \frac{Q_p(\tg_1+\tg_2)^2 \eta^2}{hf} (\cos^2\theta d\tpsi+\sin^2\theta d\tphi)^2 \nn\\
 &&-2\frac{\sqrt{Q_1 Q_5}}{hf} \left(\tg_1 \cos^2\theta d\tpsi +\tg_2 \sin^2\theta d\tphi\right) (dt-dy)\nn\\
 &&-2\frac{(\tg_1+\tg_2) \eta \sqrt{Q_1 Q_5}}{hf} \left(\cos^2\theta d\tpsi +\sin^2\theta d\tphi\right)dy\,,
\end{eqnarray}
where
\begin{equation}
\eta\equiv \frac{Q_1 Q_5}{Q_1 Q_5+Q_1 Q_p+Q_5 Q_p}\,,
\end{equation}
\begin{equation}
 f=r^2+(\tg_1+\tg_2) \eta (\tg_1 \sin^2\theta + \tg_2 \cos^2\theta)\,,
\end{equation}
\begin{equation}
\tg_1 =a \frac{j_\psi}{n_1 n_5}=-a n\,, \quad \tg_2=a\frac{j_\phi}{n_1 n_5}=a(n+1)\,,
\end{equation}
while the functions $h,H_1,H_5$ are as in~\S\ref{sec:2charge}. 
The dilaton and gauge fields are
\begin{eqnarray}
\label{3chargedilrr}
e^{\Phi} &=& g\sqrt{\frac{H_1}{H_5}},\\
C_2 &=& -\frac{\sqrt{Q_1 Q_5}\cos^2\theta}{H_1 f}\left(\tg_2 dt + \tg_1 dy\right)\wedge d\psi +\frac{\sqrt{Q_1 Q_5}\cos^2\theta}{H_1 f}\left(\tg_1 dt + \tg_2 dy\right)\wedge d\phi\nonumber\\
&&+ \frac{\left(\tg_1 + \tg_2\right)\eta Q_p}{\sqrt{Q_1 Q_5} H_1 f}\left(Q_1 dt + Q_5 dy\right) \wedge \left( \cos^2\theta d\psi + \sin^2\theta d\phi\right)\nonumber\\
&&- \frac{Q_1}{H_1 f} dt\wedge dy - \frac{Q_5 \cos^2\theta}{H_1 f} \left(r^2+\tg_2(\tg_1+\tg_2)\eta+Q_1\right)d\psi\wedge d\phi.
\end{eqnarray}
This solution has $n_1$ units of D1 branes and $n_5$ units of D5 branes wrapping the $S^1$, $n_p$ units of momentum along the $S^1$ and $j_\psi$, $j_\phi$ units of angular momenta on the $S^3$. The dimensionful quantities in~\eqref{eq:3metricR} are related to these quantized values by (using~\eqref{eq:np})
\begin{equation}
  Q_1=\frac{g \alpha'^3}{V}n_1\,, \quad Q_5=g \alpha' n_5\,,\quad Q_p=\frac{g^2 \alpha'^4}{V R^2} n_p=-\tg_1 \tg_2\,.
  \end{equation}
For $n=0$, {\it i.e.} in the absence of momentum $Q_p=0$, we have $\eta=1\,,\tg_1=0\,, \tg_2=a$ thus recovering~\eqref{eq:metricR}.

The energy and angular momenta are
\begin{equation}
 {\mathcal{E}}=Q_1+Q_5+2Q_p\,, \quad {J}_{\tpsi}=\tg_1 R\sqrt{Q_1 Q_5} \,, \quad {J}_{\tphi}=\tg_2 R\sqrt{Q_1 Q_5} \,
\end{equation}
and the coordinate transformation correponding to spectral flow~\eqref{eq:CFTSF} with~\eqref{eq:CFTSFalpha} is given by
\begin{equation}\label{eq:geom3SF}
 \psi=\tpsi - \talpha \frac{a}{\sqrt{Q_1 Q_5}}y+(\talpha-1) \frac{a}{\sqrt{Q_1 Q_5}}t\,, \quad \phi=\tphi -\talpha \frac{a}{\sqrt{Q_1 Q_5}}t+( \talpha-1) \frac{a}{\sqrt{Q_1 Q_5}}y\,.
\end{equation}
For $\talpha=1$ this reduces to the coordinate transformation~\eqref{eq:geomSF} for which the metric in the near-horizon limit $r\ll \sqrt{Q}$ and $a\ll \sqrt{Q}\ll R$ (implying $Q_p\ll Q$ and $\eta \to 1$) becomes $AdS_3\times S^3$ dual to the NS vacuum. The exicted Ramond states obtained from spectrally flowing the NS vaccuum with~\eqref{eq:CFTSFalpha} are dual to geometries obtained from $AdS_3\times S^3$ {\it via} the coordinate transformation~\eqref{eq:geom3SF} with $\talpha=n$.

\subsection{Aichelburg-Sexl in excited $AdS_3 \times S^3$}
The same procedure as in ~\S\ref{ssec:ASinAdS} suggests that the addition of massless particles to the class of 3-charge solutions~\eqref{eq:3metricR} is described by the geometries~\footnote{ The dilaton is as in \eqref{3chargedilrr}, while the RR 2-form picks up an extra piece proportionl to $(1-\xi)$ relative to \eqref{3chargedilrr} as in \cite{Lunin:2002bj}. }
 \begin{eqnarray}\label{eq:AS3metricR}
  ds^2_R&=&-\frac{1}{\bar{h}}(dt^2-dy^2) +\frac{Q_p}{\bar{h} \bar{f}} (dt-dy)^2 +\bar{h}\bar{f} \left(\frac{d\bar{r}^2}{\bar{r}^2+(\bar{\tg}_1+\bar{\tg}_2)^2 \eta}+d\theta^2\right)\nn\\
  &&+\bar{h}\left(\bar{r}^2+\bar{\tg}_1(\bar{\tg}_1+\bar{\tg}_2)\eta -\xi\frac{(\bar{\tg}_1^2-\bar{\tg}_2^2)\eta Q_1 Q_5 \cos^2\theta}{\bar{h}^2 \bar{f}^2} \right) \cos^2\theta d\tpsi^2\nn\\
  &&+\bar{h}\left(\bar{r}^2 +\bar{\tg}_2(\bar{\tg}_1+\bar{\tg}_2)\eta+ \xi \frac{(\bar{\tg}_1^2-\bar{\tg}_2^2)\eta Q_1 Q_5 \sin^2\theta }{\bar{h}^2\bar{f}^2} \right) \sin^2\theta d\tphi^2\nn\\
  &&+ \frac{Q_p(\bar{\tg}_1+\bar{\tg}_2)^2 \eta^2}{\bar{h}\bar{f}} (\cos^2\theta d\tpsi+\sin^2\theta d\tphi)^2 \\
  &&-2\xi \frac{\sqrt{Q_1 Q_5}}{\bar{h}\bar{f}} \left({\tg}_1 \cos^2\theta d\tpsi +{\tg}_2 \sin^2\theta d\tphi\right) (dt-dy)\nn\\
  &&-2\xi \frac{({\tg}_1+{\tg}_2) \eta \sqrt{Q_1 Q_5}}{\bar{h}\bar{f}} \left(\cos^2\theta d\tpsi +\sin^2\theta d\tphi\right)dy\,\nn
 \end{eqnarray}
where
\be
\bar{\tg}_i= \sqrt{\xi} \tg_i.
\ee
In particular, in the near-horizon limit this yields an Aichelburg-Sexl shockwave propagating along both angles of the $S^3$:
\begin{eqnarray}\label{eq:3chargeshockwave}
 ds^2_{NS}&=&-\left(r^2+a^2\right)\frac{dt^2}{\sqrt{Q_1 Q_5}} + r^2 \frac{dy^2}{\sqrt{Q_1 Q_5}}+\sqrt{Q_1 Q_5} \frac{dr^2}{r^2+a^2} + \sqrt{Q_1 Q_5}\left(d\theta^2+\cos^2\theta d\psi^2+\sin^2\theta d\phi^2 \right)\nn\\
 &&+\frac{q \sqrt{Q_1 Q_5}}{f} \Big\{\left[\left(r^2+a^2\right) \frac{dt}{\sqrt{Q_1 Q_5}}+\tg_1 \cos^2\theta d\psi +\tg_2 \sin^2\theta d\phi\right]^2\nn\\
 &&\quad \quad \quad \quad \quad \quad \quad  \quad \quad -\left[r^2 \frac{dy}{Q}-\tg_2 \cos^2\theta d\psi-\tg_1 \sin^2\theta d\phi \right]^2\Big\}\,.
\end{eqnarray}
We have not checked that this is a solution other than for the trivial cases $q=0$ and $q=1$. Assuming that it is, we may then again describe an adibatic limit of the ERS instability as the growth of $q$ with time.  Again, this causes the backreacted solution to shrink as a function of time, decreasing the angular momentum by a factor $\xi=1-q$ while leaving the total energy unchanged:
\begin{equation}
 \bar{\mathcal{E}}=Q_1+Q_5+2Q_p\,, \quad \bar{J}_{\tpsi}=\tg_1 R\sqrt{Q_1 Q_5} \xi\,, \quad \bar{J}_{\tphi}=\tg_2 R\sqrt{Q_1 Q_5} \xi \,.
\end{equation}
As in the 2-charge case, the solution will continue to shrink at least until we can no longer trust the ERS analysis at $\xi\sim 1/\sqrt{n_1n_5} = 1/\sqrt{N}$. In the large $N$ limit this corresponds to taking $\xi\rightarrow 0$.  Making this replacement in~\eqref{eq:AS3metricR} yields the near-horizon metric of extremal BTZ black hole with a transverse $S^3$:
\begin{equation}
 ds_{R}^2=\frac{\bar{r}^2}{\sqrt{Q_1 Q_5}} (-dt^2+dy^2)+\frac{Q_p}{\sqrt{Q_1 Q_5}}(dt-dy)^2+\sqrt{Q_1 Q_5}\left(\frac{d\bar{r}^2}{\bar{r}^2}+d\theta^2+\cos^2\theta d\tpsi^2+\sin^2\theta d\tphi^2\right)\,.
\end{equation}
This is the near-horizon limit of the 5d non-rotating D1-D5-$p$ (Strominger-Vafa) black hole \cite{Strominger:1998yg}.
Hence this preliminary analysis suggests that, as in the 2-charge microstates, the ERS instability proceeds until the 3-charge microstate is geometrically indistinguishable from the extremal BTZ black hole outside its putative horizon.

\bibliographystyle{toine}
\bibliography{references}

\end{document}